\documentclass[twocolumn,letterpaper,aps,prc,superscriptaddress,showpacs,nofootinbib,floatfix]{revtex4}

\usepackage{graphicx}	
\usepackage{amsmath}
\usepackage{hyperref}

\usepackage{xspace}	

\newcommand{\pt}{\mbox{$p_T$}\xspace}

\newcommand{\sqsntwo}{\mbox{$\sqrt{s_{_{NN}}}=200$~GeV}\xspace}
\newcommand{\pp}{\mbox{$p$$+$$p$}\xspace}

\newcommand{\auau}{\mbox{Au$+$Au}\xspace}
\newcommand{\pbpb}{\mbox{Pb$+$Pb}\xspace}

\begin{document}


\title{ Photon production from gluon mediated quark-anti-quark annihilation at confinement }

\author{Sarah Campbell}
\affiliation{Columbia University, Nevis Labs, Irvington, NY, 10533, USA}
\email[Direct correspondence to:]{sc3877@columbia.edu}
\date{\today}

\begin{abstract}
Heavy ion collisions at RHIC produce direct photons at low transverse momentum, \pt from 1-3~GeV/c, in excess of the \pp spectra scaled by the nuclear overlap factor, $T_{AA}$.  These low \pt photons have a large azimuthal anisotropy, $v_{2}$.  Theoretical models, including hydrodynamic models, struggle to quantitatively reproduce the large low \pt direct photon excess and $v_{2}$ in a self-consistent manner.  This paper presents a description of the low \pt photon flow as the result of increased photon production from soft-gluon mediated $q$-$\bar{q}$ interactions as the system becomes color-neutral.  This production mechanism will generate photons that follow constituent quark number, $n_{q}$, scaling of $v_{2}$ with an $n_{q}$ value of two for direct photons.  $\chi^{2}$ comparisons of the published PHENIX direct photon and identified particle $v_{2}$ measurements finds that $n_{q}$-scaling applied to the direct photon $v_{2}$ data prefers the value $n_{q}=1.8$ and agrees with $n_{q}=2$ within errors in most cases.  The 0-20\% and 20-40\% \auau direct photon data are compared to a coalescence-like Monte Carlo simulation that calculates the direct photon $v_{2}$ while describing the shape of the direct photon \pt spectra in a consistent manner.  The simulation, while systematically low compared to the data, is in agreement with the \auau measurement at \pt less than 3~GeV/c in both centrality bins.  Furthermore, this production mechanism predicts that higher order flow harmonics, $v_{n}$, in direct photons will follow the modified $n_{q}$-scaling laws seen in identified hadron $v_{n}$ with an $n_{q}$ value of two.
\end{abstract}

\pacs{25.75.Dw}

\maketitle

\section{Introduction\label{Sec:Intro}}
\begin{figure*}[htbp]
    \begin{center}
        \includegraphics[width=12.cm]{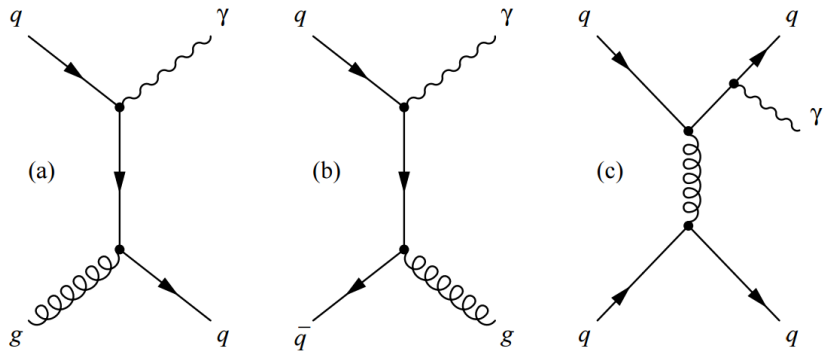}
        \caption{Feynman diagrams of prompt photon production by a) quark-gluon Compton scattering, b) quark-anti-quark annihilation, and c) bremsstrahlung radiation off of an outgoing quark~\cite{CKBThesis}.}
        \label{Fig:LOphotprod}
    \end{center}
\end{figure*}

Direct photons are all of the photons produced in a collision excluding the products of hadronic decays.  They are emitted throughout the evolution of the heavy ion medium, and because they are color-neutral they do not experience subsequent interactions with the medium.  As a result, their spectrum provides a time-integrated picture of photon emission.  Direct photons have various sources, including prompt photons generated by early hard parton interactions, photons produced in the pre-equilibrium stage, and thermal photons radiated from either the quark gluon plasma (QGP) or the hadron gas stage (HG).  In Figure~\ref{Fig:LOphotprod}, Feynman diagrams of prompt photon production mechanisms, quark-gluon Compton scattering, quark-anti-quark annihilation, and bremsstrahlung radiation, are shown.   Prompt photons are created in \pp collisions and dominate the yield at high \pt in heavy ion collisions.  Prompt photon production rates can be calculated using perturbative QCD (pQCD); quark-gluon Compton scattering and quark-anti-quark annihilation have production rates of order $\alpha_{S}\alpha$ and bremsstrahlung radiation has a rate of order $\alpha_{S}^{2}\alpha$.  QCD thermal photons have the same production diagrams, shown in Figure~\ref{Fig:LOphotprod}, but with the partons thermalized in the medium.  In thermal photon pQCD calculations, bremsstrahlung radiation is of order $\alpha_{S}\alpha$ and can exceed the production from the Compton scattering and annihilation processes.  HG thermal photons have analogous production mechanisms to the Compton scattering and annihilation processes only with pions and $\rho$-mesons interacting instead of quarks and gluons.  However, the production rates for thermal photons and other direct photons sources are not well constrained particularly in the non-perturbative regime.  This makes separating the contributions of direct photons at low and intermediate \pt difficult.

The PHENIX experiment discovered a large direct photon excess at low \pt, from 1-3~GeV/c, in \sqsntwo \auau collisions at RHIC relative to the yields of direct photons in \pp collisions scaled by the nuclear overlap factor, $T_{AA}$~\cite{ppg86,ppg162}.  
Subsequent analyses found that these low \pt photons, again from 1-3~GeV/c, have a large azimuthal anisotropy with respect to the collision's event plane~\cite{ppg126}.  Preliminary results from the ALICE experiment at the LHC suggest similar behavior in 2.76~TeV \pbpb collisions~\cite{Alicept,Alicev2}.  Hydrodynamic models are able to describe the direct photon yield with initial temperatures of 300-600~MeV and thermalization times between 0.15-0.5~fm/c~\cite{ppg86}.  Reproducing the large measured azimuthal anisotropies, $v_{2}$, at these early times has proven difficult for hydrodynamic models~\cite{Heinz,Chatterjee,Liu}.  This is because the large azimuthal anisotropies generated by hydrodynamic pressure gradients need time to develop.  To address this puzzle some theories introduce delayed QGP formation~\cite{Liu2}, new sources of photon production involving strong magnetic fields~\cite{Muller,Tuchin} and initial state Glasma effects~\cite{Glasma}, while others consider increased contributions from the hadron gas stage due to baryon-baryon and meson-baryon interactions~\cite{Rapp,Linnyk}.

In this paper, the sources of identified hadron azimuthal anisotropies are considered to understand the origin of the similarly-sized direct photon $v_{2}$.  At low \pt, bulk expansion dominates the hadronic $v_{2}$ while at high \pt, hadrons from jet fragmentation dominate.  In the intermediate \pt region, from 1-3~GeV/c, the measured baryon and meson $v_{2}$ values split, with baryons reaching higher values of $v_{2}$ at higher values of \pt~\cite{hadronflow}.  When the baryon and meson $v_{2}$ values are scaled by their number of constituent quarks, $n_{q}$, a uniform behavior between baryons and mesons is seen~\cite{nqscale}.  Coalescence models are able to reproduce quark number scaling by assuming that hadron production is dominated by the recombination of flowing partons.  They assume that thermalized co-moving quarks of a given \pt will coalesce into mesons and baryons with $n_{q}$-times the \pt and $n_{q}$-times the $v_{2}$ where $n_{q}=2$ for mesons and $n_{q}=3$ for baryons.  In this framework, energy-momentum conservation is maintained by the mean-field interaction resulting in soft gluon interactions with the medium~\cite{coalescence}.

\begin{figure}
    \begin{center}
        \includegraphics[width=4cm]{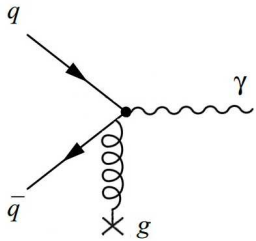}
        \caption{A Feynman diagram of the quark-anti-quark annihilation interaction with a medium gluon producing a direct photon.}
        \label{Fig:qqbar}
    \end{center}
\end{figure}

Similar mean-field or soft gluon interactions could mediate quark-anti-quark annihilation as the system moves toward color neutrality, resulting in a large increase in photon production.  These interactions, a diagram is shown in Figure~\ref{Fig:qqbar}, would produce photons from partonic processes late in the system's evolution when quarks are flowing.  One consequence of this production is that these photons should reproduce constituent quark number scaling with the value $n_{q}=2$ for direct photons.  Furthermore, this model provides a testable prediction that higher order flow harmonics, $v_{n}$, in direct photons should follow the $n_{q}$-scaling laws seen in identified hadron $v_{n}$~\cite{ppg146} again with $n_{q}=2$ for direct photons.

Section~\ref{Sec:Chi2} determines the $n_{q}$ for direct photons that best reproduces the quark number scaling seen in the identified hadron $v_{2}$ by using a $\chi^{2}$ analysis of existing data~\cite{ppg126,ppg123}.  Section~\ref{Sec:Sim} details a coalescence-like Monte Carlo calculation that combined with the $T_{AA}$-scaled \pp component is compared to the measured direct photon \pt spectrum and $v_{2}$ distribution.  A two-component model is assumed where the low \pt direct photon excess is primarily the result of quark-anti-quark annihilation mediated by mean-field or soft gluon interactions as the system becomes color neutral.

\section{The $n_{q}$-scaling of identified hadron and direct photon $v_{2}$\label{Sec:Chi2}}

\begin{figure}
    \begin{center}
        \includegraphics[width=\linewidth]{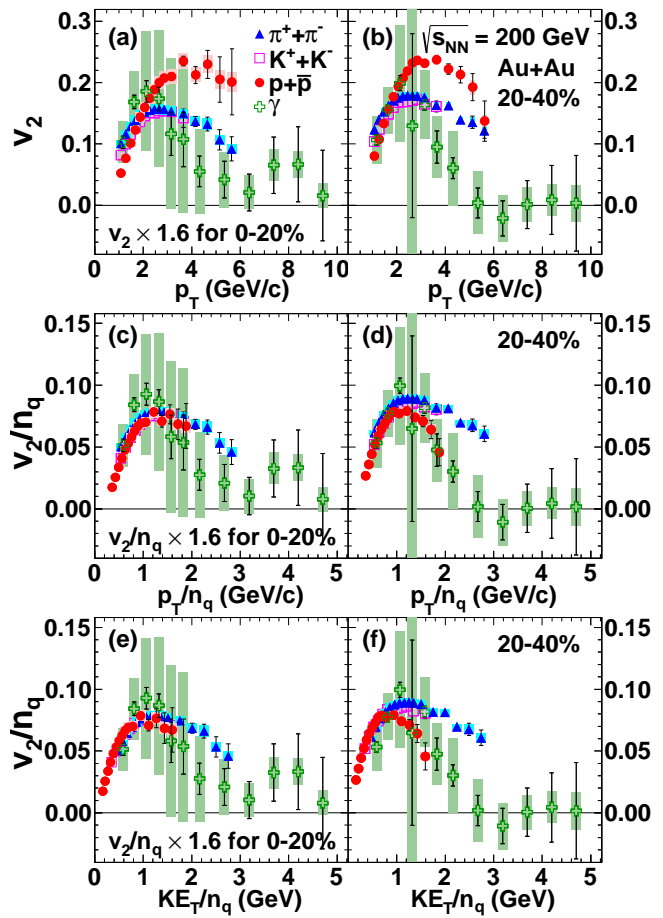}
        \caption{(color online)  The $\pi$~(blue triangles), K~(open magenta squares), p~(red circles) and direct photon~(open green crosses) $v_{2}$ as a function of \pt in central 0-20\%~(a) and mid-central 20-40\%~(b) \auau collisions at \sqsntwo.  Panels~(c) and~(d) show the $v_{2}/n_{q}$ as functions of \pt/$n_{q}$ for 0-20\% and 20-40\% respectively. Panels~(e) and~(f) show the $v_{2}$ scaled by the number of constituent quarks, $n_{q}$, as a function of $KE_{T}/n_{q}$, again for 0-20\% and 20-40\% centralities.    For direct photons, $n_{q}=2$ is assumed.  In panels~(a),~(c) and~(e), the 0-20\% $v_{2}$ values are scaled by 1.6 for better comparison to the 20-40\% results.  Error bars and shaded boxes around points represent their statistical and systematic uncertainties respectively~\cite{ppg126,ppg123}.}
        \label{Fig:nqscale2} 
    \end{center}
\end{figure}

The elliptic flow of identified hadrons displays constituent quark number scaling in the 1-3~GeV/c \pt region~\cite{ppg61,ppg124}.  In the $q$-$\bar{q}$ annihilation picture of direct photon production, this $n_{q}$-scaling behavior should extend to the direct photons with $n_{q}=2$.  This is because the $n_{q}$-scaled $v_{2}$ reflects the underlying anisotropy of the quarks and therefore is common for all hadrons and photons produced from these coalescing quarks.  At high \pt, this $n_{q}$-scaling may breakdown as contributions from hard processes begin to dominate in both the direct photon and identified hadron spectra.  Figure~\ref{Fig:nqscale2} shows a comparison of the direct photon $v_{2}$~\cite{ppg126} with the charged pion, kaon and proton $v_{2}$~\cite{ppg123} in the 0-20\% and 20-40\% \sqsntwo \auau collisions.  The $n_{q}$-scaled $v_{2}$ as a function of the $n_{q}$-scaled \pt and $KE_{T}$ are also presented assuming that the $n_{q}$ value for direct photons is two.  The agreement between the scaled direct photon $v_{2}$ and the pion, kaon and proton data is impressive despite the large systematic error bars on the direct photon measurement.  The scaled pions, kaons, protons and photons agree at low $KE_{T}/n_{q}$ in both centralities.  At $KE_{T}/n_{q}$ above 1.7~GeV, the direct photon's scaled $v_{2}$ drops below the pion values.  This deviation can be understood as the result of the increased photon production by initial hard processes~\cite{ppg126}.  Of particular note is how the direct photon and proton $v_{2}/n_{q}$ track together as they deviate from the pion values in the 20-40\% centrality bin.  This suggests a similar transition to the high \pt hard scattering region for the scaled protons and photons.  While the 0-20\% proton $v_{2}$ does not extend high enough in $KE_{T}/n_{q}$, protons in the 0-20\% centrality are also expected to break $n_{q}$-scaling at high $KE_{T}/n_{q}$ and deviations are seen in the 10-20\% bin~\cite{ppg123}.

A $\chi^{2}$ analysis is undertaken to determine if $n_{q}=2$ best produces the agreement between the direct photon and the $n_{q}$-scaled identified hadron $v_{2}$ data.  This is done in two ways.  In Section~\ref{Sec:Chi2data}, the datasets are compared directly.  In Section~\ref{Sec:Chi2fit}, the $n_{q}$-scaled identified hadron $v_{2}$ are fit and the direct photon $v_{2}$ are compared to that function.

\subsection{$\chi^2$ comparison between the direct photon and $n_{q}$-scaled hadron data\label{Sec:Chi2data}}

\begin{figure*}[htbp]
\begin{center}
\includegraphics*[width=\linewidth]{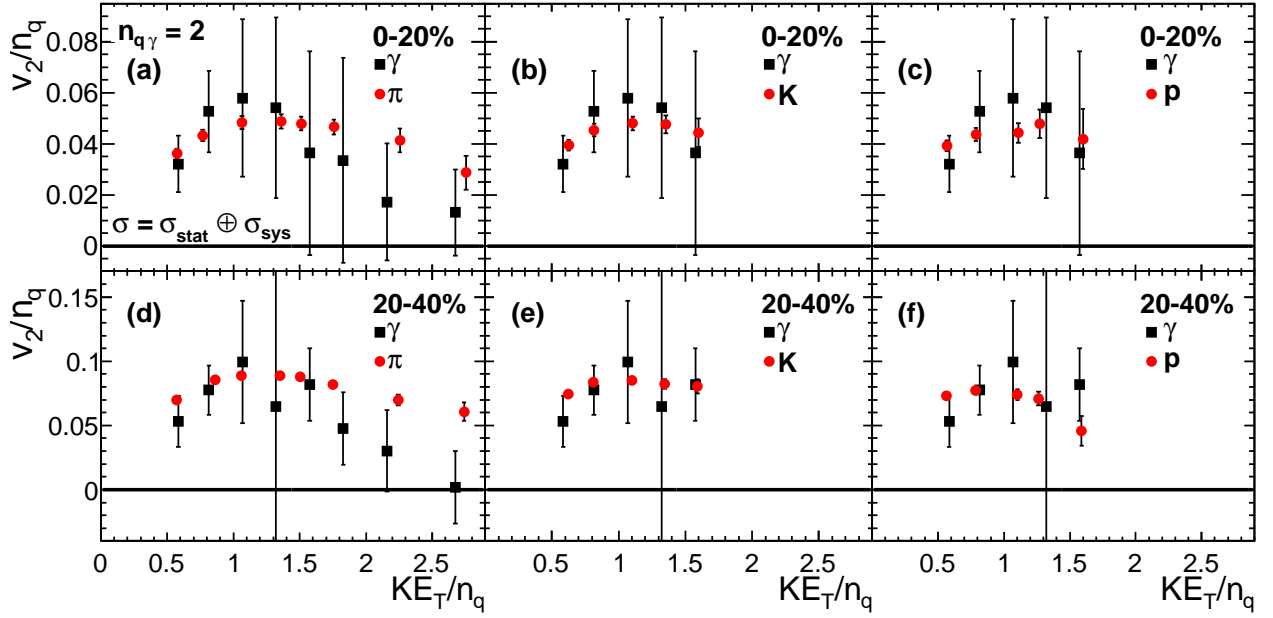}
\caption{(color online)  Example plots of the input data used for the calculation of $\chi^{2}$ comparing the $v_{2}/n_{q}$ vs $KE_{T}/n_{q}$ for identified hadrons~(red circles)~\cite{ppg123} and direct photons~(black squares)~\cite{ppg126} using the quadratic sum of the statistical and systematic errors.  Here, $n_{q\gamma}=2$ is assumed for direct photons.  The 0-20\%~(top row) and 20-40\%~(bottom row) \sqsntwo \auau results are shown.  Pions~(left column), kaons ~(middle column) and protons~(right column) are separately plotted with the direct photon data over the full $KE_{T}/n_{q}$ range.  The data are included in the $\chi^{2}$ calculation only if the identified hadron and direct photon $KE_{T}/n_{q}$ values are within 0.1~GeV/c.  The $\chi^{2}$ is calculated using the variation between direct photon and identified hadron $v_{2}/n_{q}$ in all six plots.  Error bars represent the statistical and systematic uncertainties summed in quadrature.  A $\chi^{2}/NDF$ of $16.28/35=0.47$ is found using the full $KE_{T}/n_{q}$ range available in the data. }
\label{Fig:Chi2Calc}
\end{center}
\end{figure*}

\begin{figure*}[htbp]
\begin{center}
  \includegraphics[width=0.49\linewidth]{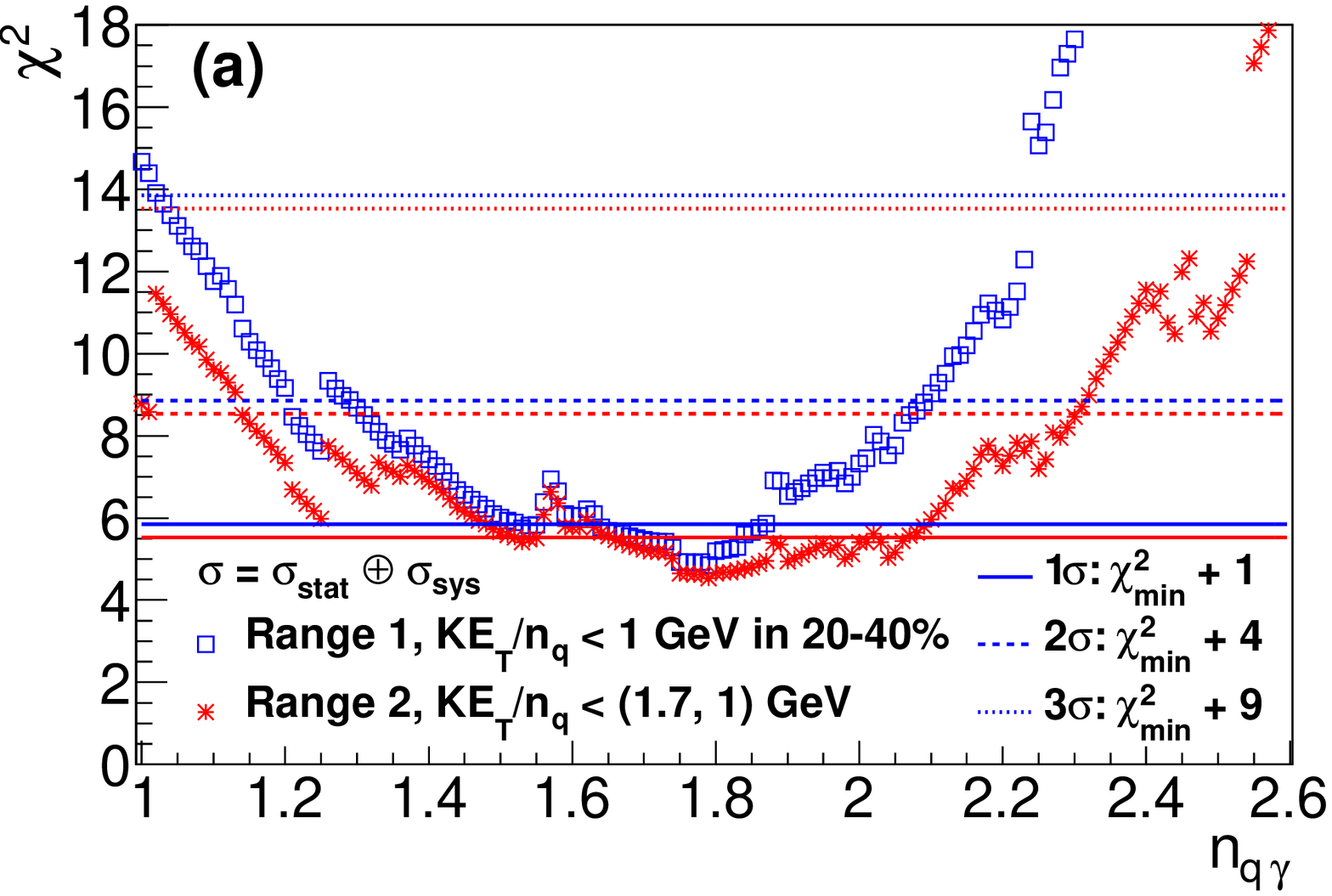}
  \includegraphics[width=0.49\linewidth]{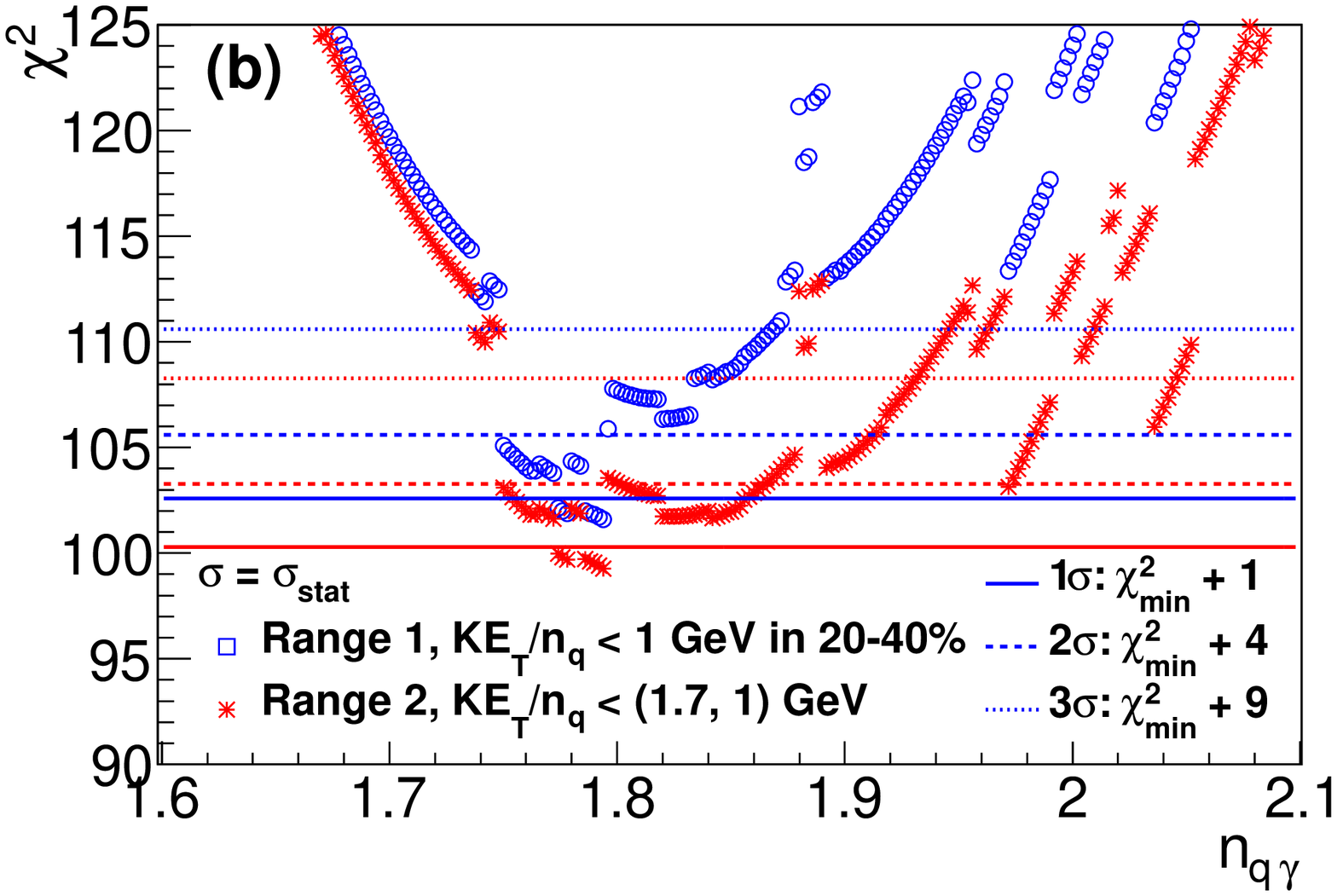}
  \caption{(color online)  The $\chi^{2}$ distribution as a function of $n_{q\gamma}$ calculated using the quadrature sum of the statistical and systematic errors for the hadron and photon uncertainties~(a) and calculated using only the statistical errors~(b).  The $\chi^{2}$ calculation with an upper limit of $KE_{T}/n_{q} < 1.0$~GeV in the 20-40\% centrality bin is shown with blue open circles; this is Range 1.  The calculation with upper limits of 1.7 and 1.0~GeV in the 0-20\% and 20-40\% centrality bins respectively is shown with red $*$ marks; this is Range 2.  Horizontal lines are drawn at the location of the $\chi^{2}_{min} + 1$~(solid), $\chi^{2}_{min}+4$~(dashed) and $\chi^{2}_{min}+9$~(dotted) for each calculation.}
    \label{Fig:Chi2Results}
\end{center}
\end{figure*}

A $\chi^{2}$ comparison is performed between the $v_{2}$ for direct photons to the $n_{q}$-scaled hadron data.  The $\chi^{2}$ comparison of the direct photon and identified hadron data is calculated according to
\begin{equation}
\chi^{2} = \sum\limits_{Cent.} \sum\limits_{\pi,K,p} \sum\limits_{KE_{T}/n_{q}} \frac{\left( v_{2\gamma}/n_{q\gamma}- v_{2h}/n_{q}\right)^2}{(\sigma_{\gamma}/n_{q\gamma})^2 + (\sigma_{h}/n_{q})^2}
\label{Eq:Chi2data}
\end{equation}
where $v_{2\gamma}$ is the direct photon $v_{2}$, $v_{2h}$ is the identified hadron $v_{2}$ for each of the summed hadrons, $\pi$, K and p.  The $\chi^{2}$ is summed over the 0-20\% and 20-40\% centralities comparing the $n_{q}$-scaled pion, kaon and proton $v_{2}/n_{q}$ values to the direct photon $v_{2}$/$n_{q\gamma}$ where $n_{q\gamma}$ is the only parameter.  Determining the photon and hadron uncertainties, $\sigma_{\gamma}$ and $\sigma_{h}$, is complicated because the published systematic errors for both the direct photons and identified hadrons combine both point-to-point and correlated systematic errors~\cite{ppg126,ppg123}.  To address this the $\chi^{2}$ analysis is performed in two ways.  In one case, the quadrature sum of the statistical and systematic errors for direct photons and the identified hadron uncertainties is used, $\sigma = \sigma_{stat} \oplus \sigma_{sys}$.  This assumes that the systematic errors are uncorrelated.  Another $\chi^{2}$ analysis assumes that the systematic errors are fully correlated and the photon and hadron uncertainties are limited to their statistical errors, $\sigma = \sigma_{stat}$.  In both cases, the comparison of a given pair of direct photon and hadron data points are included in the $\chi^{2}$ calculation only if the $KE_{T}/n_{q}$ values are within 0.1~GeV/c of each other.  An example of this data comparison over the full range in $KE_{T}/n_{q}$ is shown in Figure~\ref{Fig:Chi2Calc} where the photon-to-identified hadron data comparison plots with $n_{q\gamma}=2$ are presented.  A $\chi^2$ of $16.28$ is calculated using the quadrature sum of the statistical and systematic errors for the photon and hadron uncertainties with $35$ degrees of freedom, $NDF$, and a reduced $\chi^2$, $\chi^{2}/NDF$, of $0.47$ is found.  As a result of requiring photon-hadron matching in $KE_{T}/n_{q}$, the number of degrees of freedom of the $\chi^{2}$ calculation changes as $n_{q\gamma}$ varies.  This leads to a discontinuous $\chi^{2}$ distribution as a function of $n_{q\gamma}$, as seen in Figure~\ref{Fig:Chi2Results}.

Figure~\ref{Fig:Chi2Results}~(a) shows the $\chi^{2}$ versus $n_{q\gamma}$ when statistical and systematic errors are used to determine the $\chi^{2}$ and Figure~\ref{Fig:Chi2Results}~(b) shows the $\chi^{2}$ when only statistical errors are included.  Open circles identify the $\chi^{2}$ values when an upper limit of $KE_{T}/n_{q} < 1$~GeV is applied in the 20-40\% centrality bin.  This is Range 1.  It removes the region where the proton and pions deviate from $n_{q}$-scaling~\cite{ppg123}.  Another $\chi^{2}$ comparison, shown with $*$ marks and referred to as Range 2, restricts the $KE_{T}/n_{q}$ range in both centrality bins with upper limits of 1.7~GeV and 1.0~GeV for the 0-20\% and 20-40\% centralities respectively.  This extends the $KE_{T}/n_{q}$ cut to central collisions where the $n_{q}$-scaling is expected to remain broken~\cite{ppg123}. When the $KE_{T}/n_{q}$ range is restricted the width of the $\chi^2$ distribution increases reflecting the reduced resolving power of the $\chi^{2}$ comparison when fewer data points are included.

The optimal $n_{q\gamma}$ values for $n_{q}$-scaling are located at the $\chi^{2}$ minima, a value of 1.79 for all four $\chi^{2}$ data comparisons.  The error on the $n_{q\gamma}$ parameter is related to the width of the $\chi^{2}$ curve.  It is determined from the range of $n_{q\gamma}$ values where the $\chi^{2}$ is below $\chi_{min}^{2} + 1$ for the $1\sigma$ limit, $\chi_{min}^{2}+4$ for the $2\sigma$ limit, and $\chi_{min}^{2}+9$ for the $3\sigma$ limit.  Horizontal lines are drawn at the $\chi^{2}_{min}+n$ values in Figure~\ref{Fig:Chi2Results} with solid lines for the $1\sigma$ limits, dashed lines for the $2\sigma$ limits and dotted lines for the $3\sigma$ limits.  When the systematic errors are assumed to be fully correlated, the $\sigma=\sigma_{stat}$ case, the $n_{q}$'s systematic error from the correlation must also be obtained.  The systematic error on the $n_{q\gamma}$ in the $\sigma = \sigma_{stat}$ case is found by shifting all of the photon and identified hadron $v_{2}$ values to the extreme maximum or minimum values in their systematic error ranges, re-calculating the $\chi^{2}$ in the $n_{q\gamma}$-space, and determining the $n_{q\gamma}$ where $\chi^{2}$ reaches a minimum value.  The optimal $n_{q\gamma}$ values and errors from this comparison of data points are shown with their respective $\chi^{2}/NDF$ in Table~\ref{Tab:Chi2Results}.

\subsection{$\chi^2$ analysis using fit to $n_{q}$-scaled hadron data\label{Sec:Chi2fit}}

Here, a fit to the $n_{q}$-scaled identified hadron data is used to describe the universal scaling distribution.  The 0-20\% and 20-40\% direct photon data is then compared to this function and fit using TMinuit to find the optimal $n_{q\gamma}$ by minimize the $\chi^{2}$,
\begin{equation}
\chi^{2} = \sum\limits_{Cent.} \sum\limits_{KE_{T}/n_{q}} \frac{\left( v_{2\gamma}/n_{q\gamma}- v_{2fit}\right)^2}{(\sigma_{\gamma}/n_{q})^2}
\label{Eq:Chi2}
\end{equation}
where $v_{2\gamma}$ is the direct photon $v_{2}$ and $v_{2fit}$ is the fit to the $n_{q}$-scaled identified hadron $v_{2}$.  The $\chi^{2}$ is summed over the 0-20\% and 20-40\% centralities comparing the $v_{2fit}$ to the direct photon $v_{2}$/$n_{q\gamma}$ where $n_{q\gamma}$ is the only parameter.  Again, the $\chi^{2}$ minimization is performed in two cases to address how the direct photon uncertainty, $\sigma_{\gamma}$, relates to the direct photon systematic errors.  One case uses the quadrature sum of the statistical and systematic errors for direct photons, $\sigma = \sigma_{stat} \oplus \sigma_{sys}$.  This assumes the systematic errors are uncorrelated.  The second case assumes that the systematic errors are fully correlated and the photon uncertainties are limited to the statistical errors, $\sigma=\sigma_{stat}$.

\begin{figure}
\begin{center}
\includegraphics*[width=\linewidth]{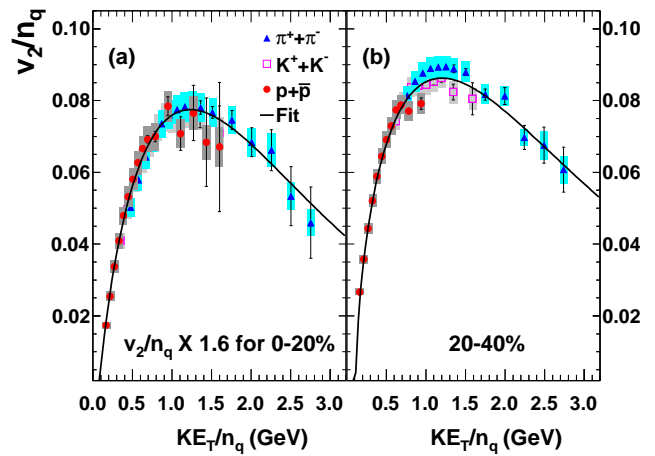}
\caption{(color online) The 0-20\% and 20-40\% \auau $v_{2}/n_{q}$ vs $KE_{T}/n_{q}$ for pions, kaons and protons are fit with a probability density distribution of a Gamma function.  High \pt protons that deviate from $n_{q}$-scaling in the 20-40\% centrality bin are excluded from the fit and are not shown~\cite{ppg123}.}
\label{Fig:v2Fit}
\end{center}
\end{figure}

To obtain $v_{2fit}$, the $n_{q}$-scaled identified hadron data is fit using a scaled probability density function of the gamma distribution,
\begin{equation}
G(x) = A \frac{\left(\left(x-\mu\right)/\beta\right)^{\gamma-1}e^{-1\left(x-\mu\right)/\beta}}{\beta \Gamma\left(\gamma\right)}
\label{Eq:GammaDist}
\end{equation}
where $x$ is $KE_{T}/n_{q}$, $\gamma$ is the shape parameter, $\mu$ is the location parameter, $\beta$ is the scale parameter, $A$ is an overall normalization scale and $\Gamma\left(\gamma\right)$ is the gamma distribution, $\Gamma\left(x\right) = \int_{0}^{\infty} t^{x-1}e^{-t}dt$.  Figure~\ref{Fig:v2Fit} shows the fit results when the 0-20\% and 20-40\% \auau identified hadron $v_{2}/n_{q}$ data are fit to Equation~\ref{Eq:GammaDist}.  In the 20-40\% centrality bin, high $KE_{T}/n_{q}$ protons that deviate from the $n_{q}$-scaled pions are excluded from the fit and are not shown.  Table~\ref{Tab:Params} lists the parameters obtained from the fits for both centrality bins.

\begin{table}[h!]
\caption{Table of the results of a Gamma distribution fit to the \auau $v_{2}/n_{q}$ vs $KE_{T}/n_{q}$.}
\label{Tab:Params}
\begin{ruledtabular}
\begin{tabular}{lcc} \hline
Parameters   &  0-20\% & 20-40\% \\ \hline
$\gamma$     &    1.86 &    1.62 \\ 
$\mu$        &    0.08 &    0.11 \\ 
$\beta$      &    1.34 &    1.76 \\ 
A            &   0.166 &    0.34 \\
\end{tabular}
\end{ruledtabular}
\end{table}

A TMinuit fit is used to determine the $n_{q\gamma}$ where the $\chi^{2}$ from Equation~\ref{Eq:Chi2} reaches its minimum value.  This fit is performed over two ranges.  Range 1 removes the region where the proton breaks the $n_{q}$-scaling~\cite{ppg123} by applying an upper limit at $KE_{T}/n_{q} < 1$~GeV in the 20-40\% centrality bin.  Range 2 is restricts the $KE_{T}/n_{q}$ range in both centrality bins with upper limits of 1.7~GeV and 1.0~GeV for the 0-20\% and 20-40\% centralities respectively.  This removes the region in the 0-20\% bin where $n_{q}$-scaling is expected to be broken~\cite{ppg123}.  TMinuit finds the optimal $n_{q\gamma}$ value with statistical errors.  When the direct photon systematic errors are assumed to be fully correlated, the $\sigma=\sigma_{stat}$ case, the $n_{q\gamma}$'s systematic errors from this correlation must also be determined.  This is done by shifting the direct photon $v_{2}$ values to the extreme maximum and minimum of the systematic error range and re-fitting with TMinuit to find $n_{q\gamma}$ at the $\chi^{2}$ minimum value.  The resulting $n_{q\gamma}$ values and errors from the TMinuit fits are shown in Table~\ref{Tab:Chi2Results} with their respective $\chi^{2}/NDF$.

\begin{table*} 
\caption{Table of optimal $n_{q\gamma}$ values and errors with $\chi^{2}/NDF$}
\label{Tab:Chi2Results}
\begin{ruledtabular}
\begin{tabular}{lcccc} 
&\multicolumn{2}{c}{$\sigma_{\gamma}=\sigma_{stat}\oplus\sigma_{sys}$}&\multicolumn{2}{c}{$\sigma_{\gamma}=\sigma_{stat}$}\\
                  & $n_{q\gamma}\pm(stat)$    & $\chi^{2}/NDF$ & $n_{q\gamma}\pm(stat)\pm(sys)$  & $\chi^{2}/NDF$ \\ \hline
Data, Range 1     & $1.79_{-0.27}^{+0.08}$ & $ 4.85/20 = 0.24$ & $1.79_{-0.01-0.72}^{+0.002+0.67}$ & $101.6/20 = 5.1$ \\
Data, Range 2     & $1.79 \pm 0.27$ & $ 4.53/17 = 0.27$ & $1.79_{-0.01-0.72}^{+0.002+1.09}$ &  $99.5/17 = 5.9$ \\
Fit, Range 1      & $1.59 \pm 0.22$ &  $3.51/13 = 0.26$ & $1.79 \pm 0.02_{-0.68}^{+0.85}$ &  $44.67/14 = 3.19$ \\
Fit, Range 2      & $1.83 \pm 0.44$ &  $1.55/5 = 0.31$ & $1.88 \pm 0.07_{-0.71}^{+1.18}$ &   $34.14/6 = 5.68$ \\
\end{tabular}
\end{ruledtabular}
\end{table*}

The low $\chi^{2}/NDF$ values under the $\sigma_{\gamma}=\sigma_{stat}\oplus\sigma_{sys}$ heading reflect the over-estimation of the photon and hadron uncertainties when uncorrelated systematic errors are assumed.  Under the $\sigma_{\gamma}=\sigma_{stat}$ heading, when only the statistical errors are used in the $\chi^{2}$ determination, the corresponding $\chi^{2}/NDF$ values are above one, a consequence of the underestimation of the uncertainty when the systematic errors are assumed to be fully correlated.  The separation of the systematic errors into errors that are point-to-point independent and those that are correlated is needed to fully interpret the $\chi^{2}/NDF$ values in these comparisons.

The hypothesized value of $n_{q\gamma}=2$ is within the systematic uncertainty region when the $n_{q\gamma}$ is determined from the data with $\sigma_{\gamma}=\sigma_{stat}$ in both Range 1 and Range 2.  The $n_{q\gamma}=2$ condition is inside of the $1\sigma$ limit for the $\sigma_{\gamma}=\sigma_{stat}\oplus\sigma_{sys}$, Range 2 data comparison and within the $2\sigma$ limit for the $\sigma_{\gamma}=\sigma_{stat}\oplus\sigma_{sys}$, Range 1 data comparison.  The $n_{q\gamma}$ values from the comparison to the fit of the $n_{q}$-scaled hadron data are very similar to the direct data comparison results. A $n_{q\gamma}$ value close to 1.8 is found over both ranges when $\sigma=\sigma_{stat}$ is assumed and in Range 2 when $\sigma=\sigma_{stat}\oplus\sigma_{sys}$ is assumed.  Only the TMinuit fit over Range 1 produces a $n_{q\gamma}$ value that differs from 1.8, however, it is within $2\sigma$ of the $n_{q\gamma}=2$ hypothesis.  Of the eight $n_{q\gamma}$ searches presented here, six are consistent with $n_{q\gamma}=2$ within $1\sigma$.  The remaining two $n_{q\gamma}$ searches are consistent with the $n_{q\gamma}=2$ hypothesis at the $2\sigma$ level.  These two comparisons both use the larger $KE_{T}/n_{q}$ region in the 0-20\% centrality and $\sigma = \sigma_{stat} \oplus \sigma_{sys}$.  These comparisons are affected by the difference between the pion $v_{2}$ and direct photon $v_{2}$ at $KE_{T}/n_{q} > 1.7$~GeV in the 0-20\% centrality bin, seen in Figure \ref{Fig:nqscale2}.  This difference between the pion and direct photon $v_{2}$ at high $KE_{T}/n_{q}$ is the result of the increased direct photon contributions from hard scattering at high $p_{T}$, $p_{T} > 3.5$~GeV~\cite{ppg126}.

The large systematic errors in the direct photon data dominate the uncertainty in the $n_{q\gamma}$ determination.  Reduced systematic errors on the direct photon $v_{2}$ measurement and separating the systematic errors into errors that are point-to-point independent and those that are correlated would reduce the uncertainty and improve the calculation of the $\chi^{2}$ in these comparisons.  Proton $v_{2}$ measurements that extend out to higher \pt in the 0-20\% centrality bin, and direct photon $v_{2}$ measurements in additional centrality bins and collision systems would provide additional points for comparison benefiting this analysis by reducing the width of the $\chi^{2}$ distribution and improving the resolving power of the $n_{q\gamma}$ parameter.  Furthermore, direct photon azimuthal anisotropy measurements at higher orders, $v_{n}$, will provide an additional tests to this model.  The model predicts that higher order direct photon $v_{n}$ will follow the higher-order modified $n_{q}$ scaling relation, with a universal curve in $v_{n}/n_{q}^{n/2}$ as a function of $KE_{T}/n_{q}$~\cite{ppg146}, with $n_{q\gamma}=2$ for direct photons.

Seven out of the eight $\chi^{2}$ comparisons shown here find an optimum $n_{q\gamma}$ value of approximately 1.8.  In six cases, the $n_{q\gamma}=2$ condition is within $1\sigma$ of the optimum value.  In the remaining two cases, the $n_{q\gamma}=2$ condition is within $2\sigma$ of the optimum value.  These two cases are biased by the hard scattering contributions at high $p_{T}$.  These results, in conjunction with the similarity in the data seen in Figure~\ref{Fig:nqscale2}, indicate that the direct photon $v_{2}$ data are consistent with the hypothesis of $n_{q\gamma}=2$ required by the $q$-$\bar{q}$ annihilation production mechanism.

\section{Simulating the direct photon $v_{2}$\label{Sec:Sim}}

To further develop the ansatz of photon production at confinement from coalescence-like quark-anti-quark annihilation, a data-driven Monte Carlo simulation is developed.  The crux of the direct photon puzzle is to reconcile the $p_{T}$ spectral shape with the large azimuthal anisotropy.  In Section~\ref{Sec:MCshape}, the $q$-$\bar{q}$ photon \pt spectral shape and $v_{2}$ are simulated with a Monte Carlo simulation.  Rather than calculating the yields, a fit to the measured \pt distribution is performed in Section~\ref{Sec:MCfit} to determine if the $q$-$\bar{q}$ photon \pt shape from the Monte Carlo is able to describe the large excess above the $T_{AA}$-scaled \pp yield seen in the data. Then the direct photon $v_{2}$ is calculated by weighting the $q$-$\bar{q}$ photon $v_{2}$ by the relative contribution of the $q$-$\bar{q}$ photon component to the total direct photon yield; the $T_{AA}$-scaled \pp contribution is assumed to be azimuthally isotropic.

\subsection{Monte Carlo of coalescence-like $q$-$\bar{q}$ photon $v_{2}$ production\label{Sec:MCshape}}

The Monte Carlo consists of randomly sampling quark $m_{T}$ values from a thermal Blast Wave distribution.  The quark flow is implemented by calculating the quark $v_{2}$ from a fit of the measured $n_{q}$-scaled identified hadron $v_{2}$ and then sampling the quark $\phi$ from the $v_{2}$-modulated $\phi$ distribution.  This process is repeated for three quarks and then co-moving requirements are applied.

The quark's $m_{T}$ is randomly sampled from a thermal Blast Wave distribution,
\begin{multline}
\frac{d^3N}{dm_{T}dyd\phi} \propto m_{T}^{2} r \cosh(y) \\ \times exp\left(\frac{p_{T}\sinh(\rho)\cos(\phi)-m_{T}\cosh(\rho)\cosh(y)}{T}\right)
\label{Eq:BW}
\end{multline}
where T is the temperature, $m_{T}=\sqrt{p_{T}^{2}+m_{q}^{2}}$ is the transverse mass, $\rho = \tanh^{-1}(\beta_{S} (r/R)^{\alpha})$ is the boost angle, and $\phi$ is the azimuthal angle with respect to the reaction plane~\cite{BlastWave}.  Further, $\beta_{S}$ is the surface velocity, $R$ is the maximum radius in the region and $m_{q}$ is the quark mass.  A $\beta_{S}$ value of 0.75 is assumed and is consistent with $\langle \beta \rangle = 0.5$ with $\alpha$ set to one.  A quark mass of 300~MeV, temperature of 106~MeV and maximum radius of 8.5~fm is used.  The parameters of the Blast Wave distribution are taken from Refs.~\cite{Lijuan} and~\cite{Lisa}.  These Blast Wave parameters characterize the $m_{T}$ distribution of the late-stage medium and therefore identical parameters are used for the \auau $0-20\%$ and $20-40\%$ centrality bins.  The $r^{2}$, $y$ and $\phi$ values that determine the Blast Wave distribution are each chosen from flat distributions; $r$ and $y$ are the quark's radius and rapidity respectively.  The quark's $y$ is chosen from $\pm0.50$ and a $\pm0.35$ rapidity cut is applied to the resulting photons.  The random choice of $\phi$ ensures that each of the successive Blast Wave distributions sample the full variation in azimuth.

Rather than using this $\phi$ for the quark's $\phi$, the thermal quark's $\phi$ is chosen from an data-driven procedure to reduce the simulation's dependence on free parameters.  This is done by using the $m_{T}$ obtained from the Blast Wave to calculate the quark azimuthal anisotropy from a fit to the measured $n_{q}$-scaled $v_{2}$ of identified hadrons shown in Figure~\ref{Fig:v2Fit}.  Once the quark's $v_{2}$, $v_{2q}$, is calculated it is used to generate a $1 + 2v_{2q}\cos(2\phi)$ probability distribution to randomly select the quark's $\phi$.  The $v_{2q}$ is calculated using a fit to the measured $n_{q}$-scaled identified hadron $v_{2}$.  A scaled probability density function of the gamma distribution, Equation~\ref{Eq:GammaDist}, is fit to the $n_{q}$-scaled identified hadron $v_{2}$ data as described in Section~\ref{Sec:Chi2fit}.

This method effectively averages the $\phi$ variation within the Blast Wave distribution while still including radial boost effects.  By choosing the $\phi$ from the $1+2v_{2}cos(2\phi)$ distribution, the measured identified hadron $v_{2}/n_{q}$ is used to guide the modeled quark's azimuthal anisotropy.  This empirical approach to describe the quark's azimuthal anisotropy keeps the number of free parameters in the model to a minimum.  One downside of this approach is that the $v_{2q}$ from the fit relies on the pion data at high $KE_{T}/n_{q}$ which has increasing contributions from non-thermal quarks either from hard processes and fragmentation or hard thermal coalescence~\cite{Hwa}.  This may underestimate the amount of quark flow at high $KE_{T}/n_{q}$.

\begin{figure}
\begin{center}
\includegraphics*[width=\linewidth]{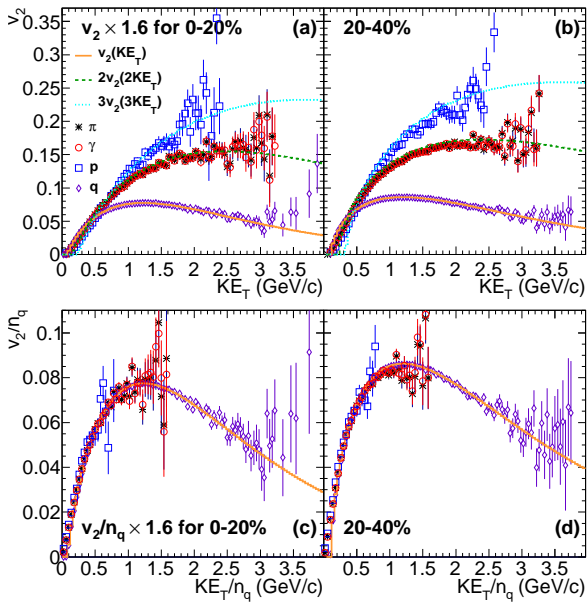}
\caption{(color online)  The $v_{2}$ for pions, photons, protons and thrown quarks simulated using the fast Monte Carlo method.  Plots~(a) and~(b) are the $v_{2}$ vs $KE_{T}$ for the 0-20\% and 20-40\% respectively.  Plots~(c) and~(d) are the $n_{q}$-scaled results for 0-20\% and 20-40\%.  The 0-20\% $v_{2}$ values are scaled by 1.6 to make the y-axis scales consistent.}
\label{Fig:MCv2}
\end{center}
\end{figure}

\begin{figure*}[htbp]
\begin{center}
  \includegraphics[width=0.49\linewidth]{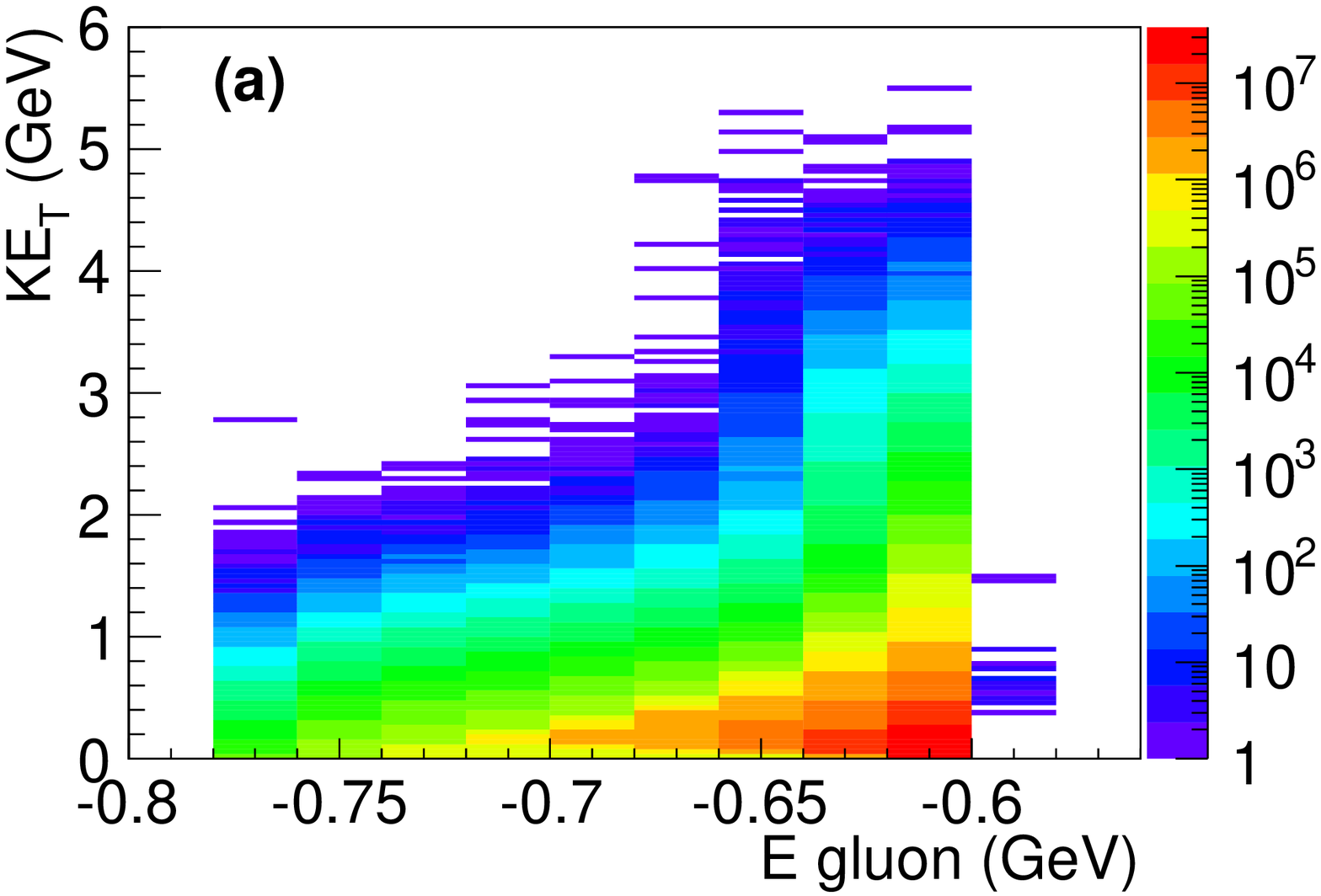}
  \includegraphics[width=0.49\linewidth]{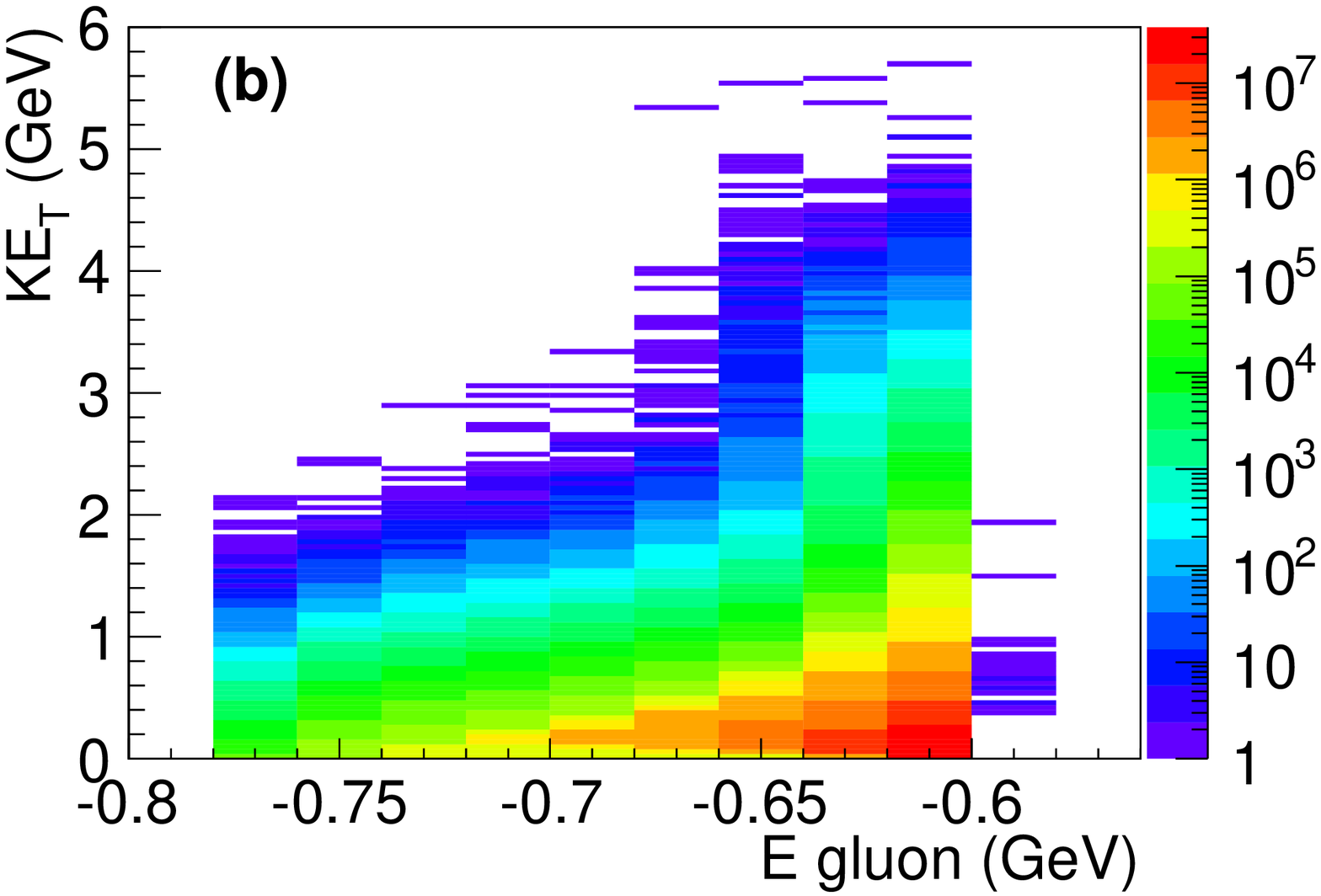}
  \caption{(color online)  The energy taken by the gluon as a function of the direct photon's $KE_{T}$ for the 0-20\%~(a) and 20-40\%~(b) simulations.  The z-axis is the number of counts and is shown with a logarithmic color scale.}
    \label{Fig:missingE}
\end{center}
\end{figure*}

The random determination of the quark's $m_{T}$ and $\phi$ is repeated for the second and third quarks within the Monte Carlo event.  The same rapidity and radius is assumed for subsequent quarks, and therefore the same Blast Wave distribution.  However, a new $m_{T}$ value is sampled, $v_{2q}$ is calculated and $\phi$ is sampled using the $1+2v_{2q}cos(2\phi)$ distribution.  The following co-moving requirements, motivated by~\cite{coalescence}, are applied to all three quarks to produce a baryon and to the first and second quarks to produce a meson,
\begin{description}
\item[Mesons:] $|p_{1} - p_{2}| < 2\Delta p$, $|x_{1} - x_{2}| < \Delta x$
\item[Baryons:] $|p_{1} - p_{2}| < \sqrt{2} \Delta p$, $|x_{1} - x_{2}| < \sqrt{2} \Delta x$, \\ $|p_{1} + p_{2} - 2p_{3}| < \sqrt{6} \Delta p$, $|x_{1} + x_{2} - 2x_{3}| < \sqrt{6} \Delta x$
\label{Eq:comoving}
\end{description}
where $p_{i}$ and $x_{i}$ are the three dimensional momentum and position vectors of the various quarks, and $\Delta p$ and $\Delta x$ are 0.2~GeV/c and 0.85~fm respectively~\cite{coalescence}.  Quarks and anti-quarks that annihilate to produce photons must satisfy the same co-moving requirements as mesons.  The four-momenta of quark pairs and triplets that satisfy the co-moving requirements are summed to create pions, photons and protons respectively. The hadrons and photons are brought on mass shell while maintaining kinetic energy conservation.  Figure~\ref{Fig:missingE} shows the amount of energy taken up by the gluon to bring the photon on mass shell as a function of the direct photon's $KE_{T}$ for the 0-20\%~(left) and 20-40\%~(right) simulations.  The z-axis is the number of counts and is shown with a logarithmic color scale.  The gluon's energy contribution is defined as $E_{\gamma} - E_{q1} - E_{q2}$ and has a value of approximately $-600$~MeV.  At photon $KE_{T}<2$~GeV, $E_{gluon}$ extends to lower energies of $-770$~MeV, however, the majority of the contribution is located at $-600$~MeV for all photon $KE_{T}$ values. This negative value means that the gluon removes some of the energy from the quarks and passes it to the medium when the photon is produced.  Additional simulations maintaining momentum conservation and energy conservation are also performed, however, kinetic energy conservation best reproduces the $n_{q}$-scaling seen in the pion and proton $v_{2}$ data.  Figure~\ref{Fig:MCv2} shows the $v_{2}$ for the thrown quarks and simulated pions, protons and photons in 0-20\%~(a) and 20-40\%~(b) centrality bins.  The $v_{2}/n_{q}$ vs $KE_{T}/n_{q}$, Figures~\ref{Fig:MCv2}~(c) and~(d), show that the $n_{q}$-scaling is well reproduced in the simulation.  Table~\ref{Tab:invslopes} displays the inverse slopes of the Monte Carlo \pt spectral shape when fit to an exponential in different \pt ranges.  These are consistent with the inverse slopes obtained from fits to the \auau data over similar \pt ranges~\cite{ppg86,ppg162}.

\begin{table}[h!]
\caption{Table of the inverse slope of the direct photon \pt spectral shape in different centralities and \pt ranges.}
\label{Tab:invslopes}
\begin{ruledtabular}
\begin{tabular}{lccc}
Centrality & \pt range            &  Monte Carlo & \auau data~\cite{ppg86,ppg162} \\ \hline
0-20\%  & $0.6$-$2.0$~GeV/c &  $233 \pm  6$ &  $239 \pm 29 \pm  7$ \\
0-20\%  & $1.0$-$2.2$~GeV/c &  $251 \pm  8$ &  $221 \pm 19 \pm 19$ \\
20-40\% & $0.6$-$2.0$~GeV/c &  $233 \pm  8$ &  $260 \pm 33 \pm  8$ \\
20-40\% & $1.0$-$2.2$~GeV/c &  $251 \pm 10$ &  $217 \pm 18 \pm 16$ \\
\end{tabular}
\end{ruledtabular}
\end{table}

\subsection{Determining the yield of the $q$-$\bar{q}$ photon component\label{Sec:MCfit}}

To find the total direct photon production, a two-component model consisting of the $q$-$\bar{q}$ photon contribution and the $T_{AA}$-scaled \pp contribution is used.  While additional photon sources are expected, these are assumed to be negligible compared to the $q$-$\bar{q}$ and $T_{AA}$-scaled \pp components.  The simulated $q$-$\bar{q}$ photon contributions are normalized to the measured direct photon yields.  The normalization constant of the $q$-$\bar{q}$ photon component is determined from a fit to the measured \auau~\cite{ppg86,ppg162,ppg139} and $T_{AA}$-scaled \pp data~\cite{ppg86,ppg136,ppg60} using TMinuit.  The normalization constant is the only parameter of the fit.  The $\chi^{2}$ is calculated using the statistical errors from the Monte Carlo simulation and the statistical and systematic errors from the data summed in quadrature.  At low \pt where \pp reference data is scarce, the \pp yield is extrapolated from the power law fit obtained from~\cite{ppg162}.  The normalization error on the $q$-$\bar{q}$ photon component and the systematic error of the $T_{AA}$-scaled \pp fit result in systematic error band on the simulation.

\begin{figure*}[htbp]
\begin{center}
  \includegraphics[width=0.49\linewidth]{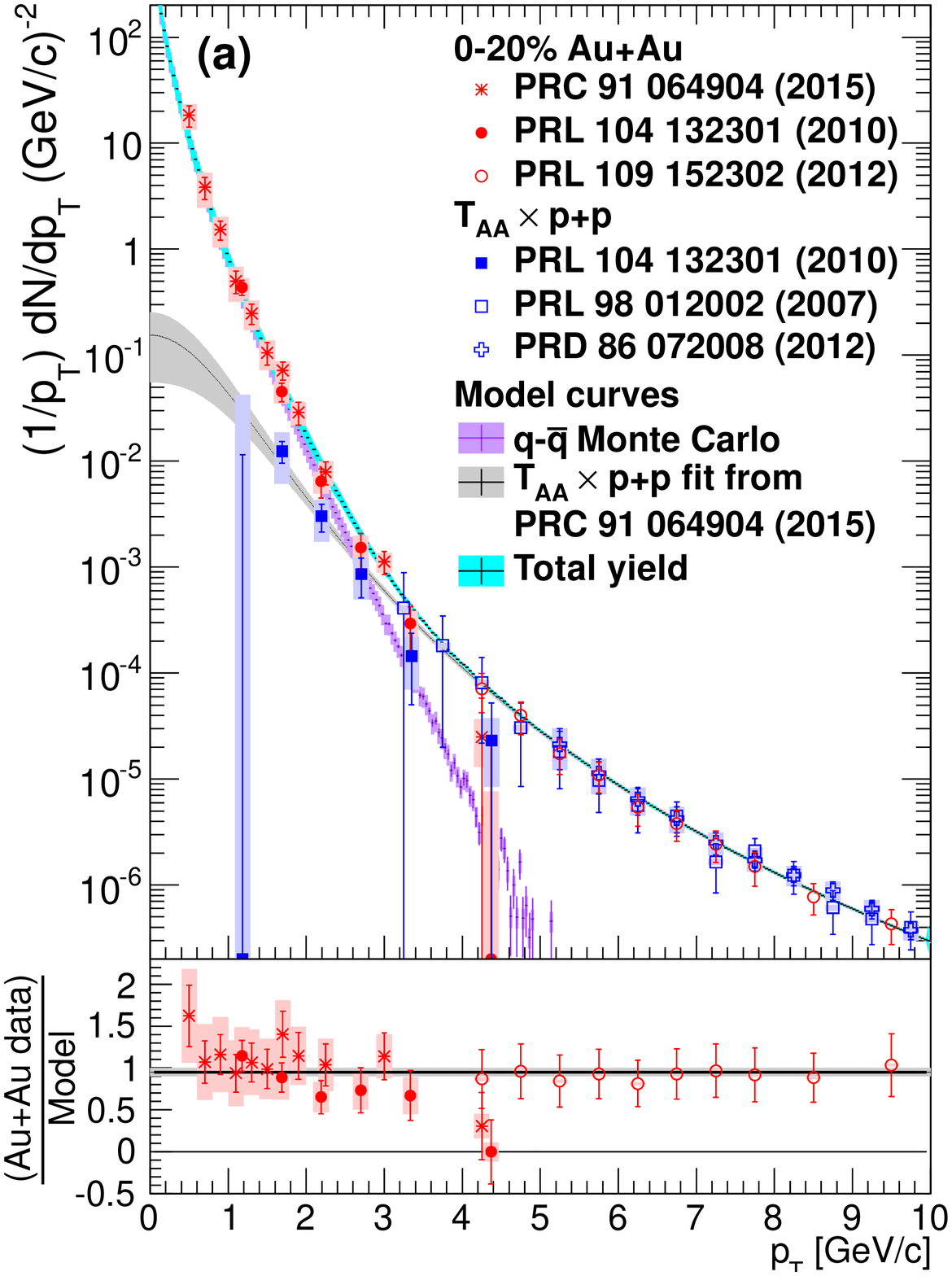}
  \includegraphics[width=0.49\linewidth]{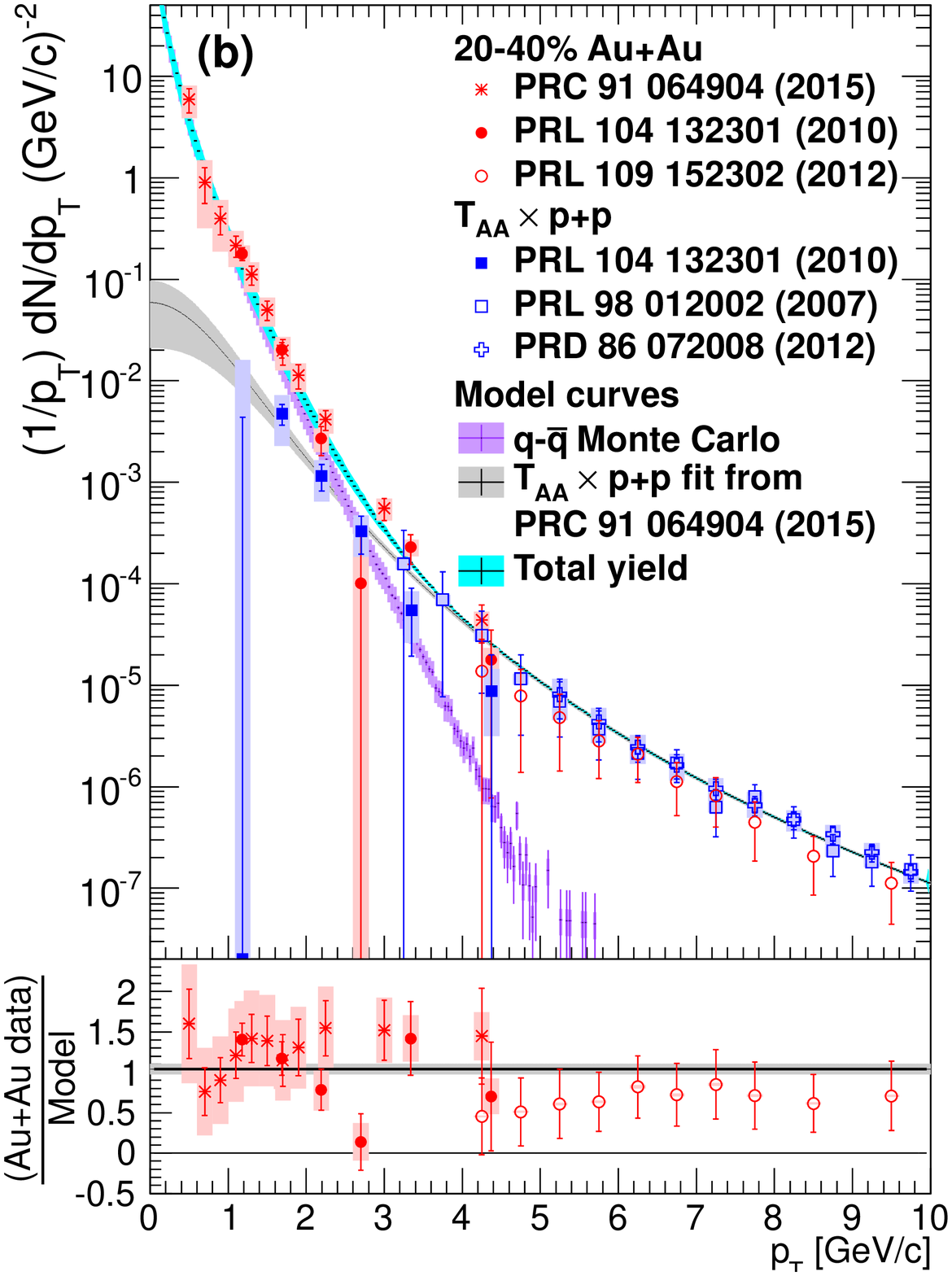}
  \caption{(color online)  The direct photon yield versus \pt for the 0-20\%~(a) and 20-40\%~(b) \auau data~(red circles and asterisks)~\cite{ppg86,ppg162,ppg139} are shown on a log-scale.  The $T_{AA}$-scaled \pp yields~(blue squares and crosses)~\cite{ppg86,ppg136,ppg60} are also shown including a power law fit to the \pp data~(grey band)~\cite{ppg162}.  The Monte Carlo yield from quark anti-quark annihilation~(purple band) are fit to the data and are shown with the total fit yield~(cyan band) found by summing the Monte Carlo yield and the $T_{AA}$-scaled \pp fit. The ratio of the \auau data over the total fit yield is shown in the lower plots.  The thick black line is a flat line fit to this ratio with a value of $0.951 \pm 0.051$ and $1.038 \pm 0.065$ for the 0-20\% and 20-40\% ratios respectively.}
    \label{Fig:Yields}
\end{center}
\end{figure*}

Figure~\ref{Fig:Yields} shows the resulting \pt distributions for 0-20\% and 20-40\% \auau collisions.  The various \auau measurements are shown in red circular symbols and the $T_{AA}$-scaled \pp measurements are shown in blue square and cross symbols.  The \pp fit is shown with a grey band, the normalized $q$-$\bar{q}$ photon contribution is shown with a purple band and the total simulated yield is shown with a cyan band.  The error on the yield determination results in a systematic band on the $q$-$\bar{q}$ photon contribution which is propagated to the total simulated yield.  Below the main figures the ratio of the \auau data to the simulation result is shown.  This ratio is fit to a flat line and found to be consistent with one for both centralities, a value of $0.951 \pm 0.051$ for 0-20\% and $1.038 \pm 0.065$ for 20-40\%.  The $\chi^{2}/NDF$ values for these flat line fits are $22.8/26 = 0.877$ and $32.5/26 = 1.25$ for the 0-20\% and 20-40\% ratios respectively. This confirms that the photons generated by the gluon-mediated annihilation of radially boosted quarks are able to describe the shape of the direct photon \pt spectra for both the 0-20\% and 20-40\% centrality bins.

\begin{figure*}[htbp]
\begin{center}
  \includegraphics[width=0.49\linewidth]{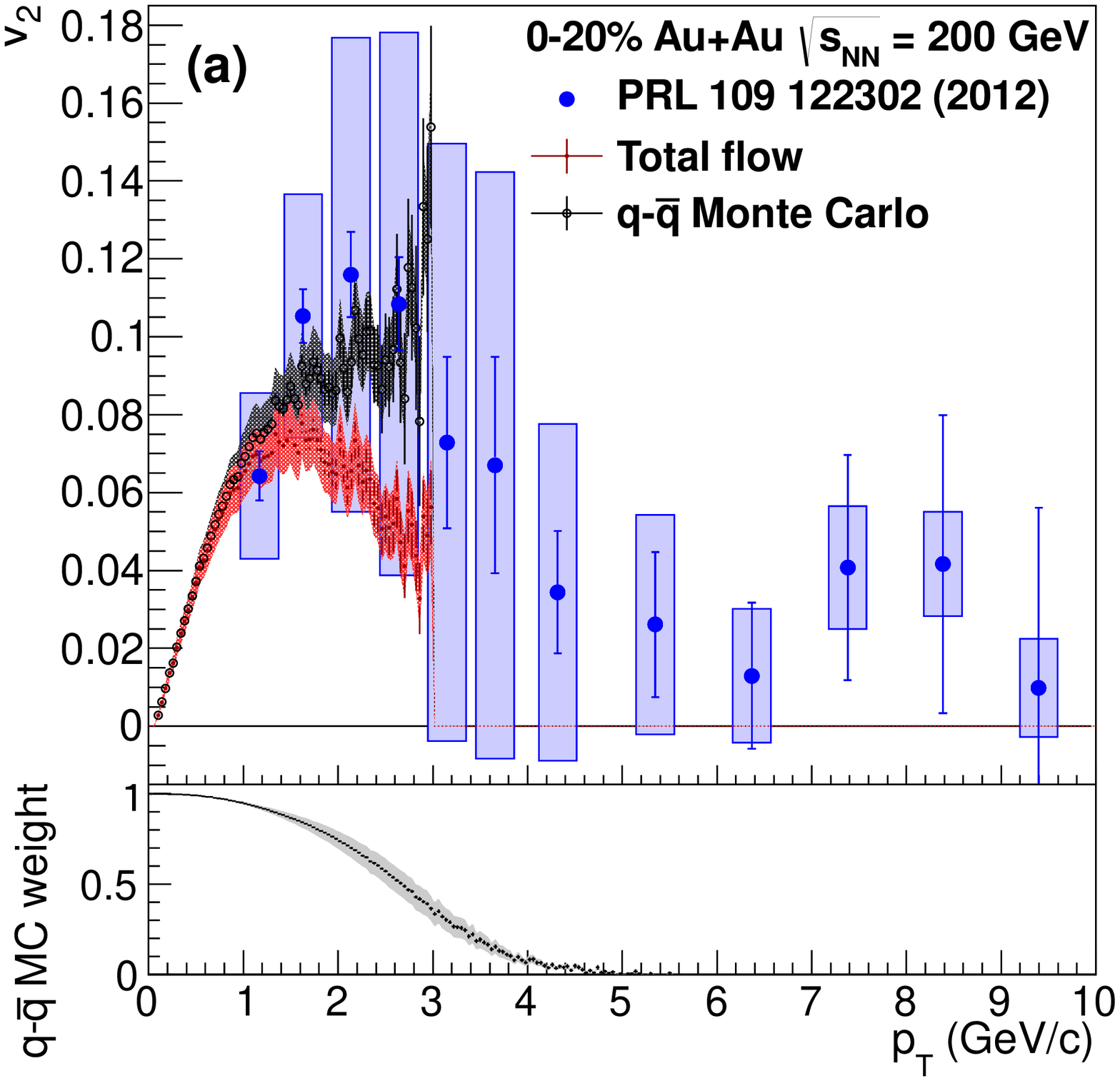}
  \includegraphics[width=0.49\linewidth]{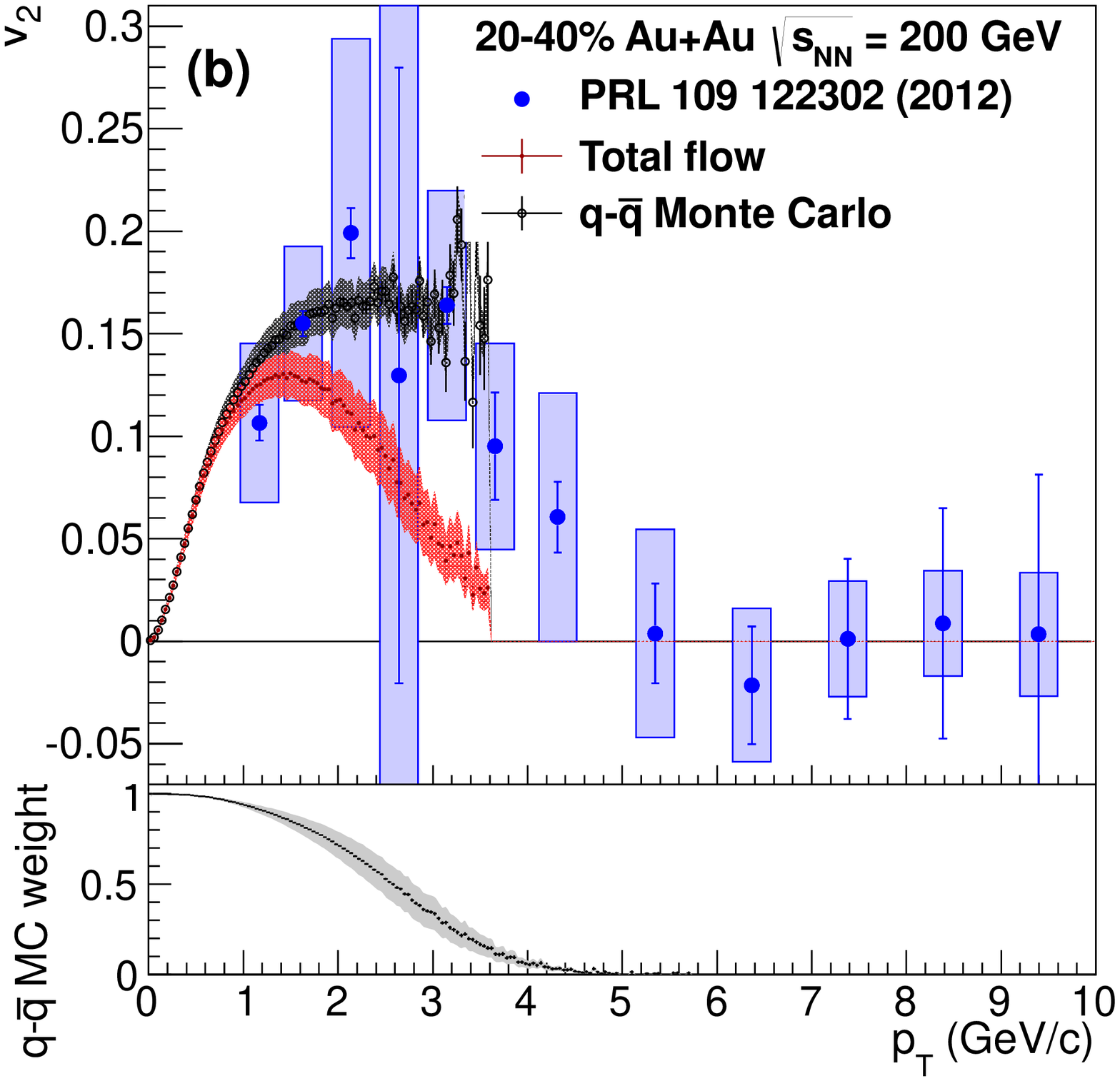}
  \caption{(color online)  The direct photon $v_{2}$ versus \pt for the 0-20\%~(a) and 20-40\%~(b) \auau data~(blue circles) is shown~\cite{ppg126}.  The Monte Carlo $v_{2}$ from quark anti-quark annihilation~(red open circles) and the total $v_{2}$~(black squares) are shown.  The relative contribution of the quark anti-quark annihilation component is shown in the lower plot of each figure.}
    \label{Fig:v2s}
\end{center}
\end{figure*}

The total direct photon $v_{2}$ is the weighted average of each component's $v_{2}$.  The $T_{AA}$-scaled \pp contribution is assumed to have no reaction plane dependence and, therefore, a $v_{2}$ of zero.  By weighting the simulated $q$-$\bar{q}$ photon $v_{2}$ by the relative contributions of the $q$-$\bar{q}$ photon yield to the total simulated yield, the total low \pt photon $v_{2}$ for each centrality can be calculated.  Figure~\ref{Fig:v2s} compares the simulated direct photon $v_{2}$ to the measured \auau $v_{2}$~(solid blue circles)~\cite{ppg126}. The open red circles are the unweighted $q$-$\bar{q}$ photon $v_{2}$ generated in the Monte Carlo.  The small black squares are the total direct photon $v_{2}$ assuming uniform azimuthal production from the $T_{AA}$-scaled \pp source.  The relative contribution of the $q$-$\bar{q}$ photon component to the yield is shown below the $v_{2}$ plots; this is the weight used to calculated the total simulated $v_{2}$.  The error in the $q$-$\bar{q}$ yield normalization lead to the systematic error in this $q$-$\bar{q}$ Monte Carlo weight.  The systematic error in the modeled $v_{2}$ is calculated from the quadrature sum of this normalization error and the systematic error on the fit to the $n_{q}$-scaled $v_{2}$ of identified hadrons, with relative error values of 10\% and 7\% in 0-20\% and 20-40\% respectively.  The model simulation of the total direct photon $v_{2}$ extends out to a \pt of 3.6~GeV/c in 0-20\% and 3.2~GeV in 20-40\%, above which the simulation lacks sufficient statistics.  For the 0-20\% centrality the total direct photon $v_{2}$ agrees with the measured results within error bars.  However, above a \pt of 1.4~GeV/c, the simulated $v_{2}$ is systematically at the bottom of the error range.  In the 20-40\% centrality comparison, the total simulated $v_{2}$ agrees with the measured results for \pt less than 3~GeV/c, above 3~GeV/c it underestimates the measured $v_{2}$.  In both centralities, the simulated direct photon $v_{2}$ agrees with the measured $v_{2}$ within errors.

\section{Conclusions}

Photon production from gluon mediated $q$-$\bar{q}$ annihilation as the system becomes color neutral is proposed as a large additional source of direct photons.  This would require direct photons follow $n_{q}$-scaling with an $n_{q\gamma}=2$.  The large direct photon flow measured in \auau collisions at RHIC is consistent with $n_{q}$-scaling when $n_{q\gamma}=2$.  Furthermore, in the 20-40\% comparison where the high \pt proton $v_{2}/n_{q}$ is seen to split from the $n_{q}$-scaled pion result, the direct photon $v_{2}/n_{q}$ follows the same trend as the proton.  This suggests that direct photons and protons may experience similar transitions from the recombination dominated intermediate \pt to the higher \pt region dominated by hard processes.  $\chi^{2}$ comparisons of the direct photon and identified hadron $v_{2}$ in $KE_{T}/n_{q}$ regions where $n_{q}$-scaling is seen in identified hadron data, find that the direct photon $v_{2}$ optimally agrees with the uniform $n_{q}$-scaled curve when $n_{q\gamma}$ is near a value of $1.8$.  Six out of eight of $\chi^{2}$ comparisons are consistent with $n_{q\gamma}=2$, the remaining two comparisons are consistent at the $2\sigma$ level.  The two remaining comparisons include the 0-20\% high $KE_{T}/n_{q}$ region where deviations from $n_{q}$-scaling are expected.  The $\chi^{2}$ comparisons would benefit from reduced systematic errors on the direct photon $v_{2}$ measurement and the separation of the systematic errors into uncorrelated and correlated errors.  Direct photon and identified hadron $v_{2}$ measurements in additional centralities, $\sqrt{s_{NN}}$ and collisions systems as well as proton $v_{2}$ measurements that extend out to higher \pt would provide further points of comparison and thus improve this analysis.

A Monte Carlo simulation generates the $q$-$\bar{q}$ annihilation photon component \pt shape and $\phi$ modulation assuming a coalescence-like framework with quarks that follow a Blast Wave $m_{T}$ distribution and a data-driven $v_{2}$ parametrization.  The Monte Carlo is able to reproduce the $n_{q}$-scaling of pions and protons and determine the $q$-$\bar{q}$ photon $v_{2}$ and the shape of its \pt distribution.  The resulting $q$-$\bar{q}$ photon \pt shape with the $T_{AA}$-scaled \pp photon yield is able to describe the large direct photon excess seen in 0-20\% and 20-40\% \auau collisions.  The simulated direct photon $v_{2}$ is consistent with the measured $v_{2}$ in the 0-20\% centrality bin but systematically low.  In the 20-40\% comparison the simulated direct photon $v_{2}$ is able to reproduce the measured direct photon $v_{2}$ at \pt less than $3$~GeV/c but underestimates the $v_{2}$ at higher \pt as the $T_{AA}$-scaled \pp contribution becomes significant.  The addition of thermal hard quark pairs would likely contribute to additional yield and flow for \pt values above $3$~GeV/c~\cite{Hwa}.  Future work would benefit from a more robust hydrodynamic calculation of the flowing quarks near the phase transition with yield estimates.  This is particularly important as the determination of the quarks $v_{2}$ from the $n_{q}$-scaled identified hadron $v_{2}$ is expected to falter at high \pt as non-thermal production mechanisms such as thermal hard coalescence and fragmentation from hard interactions contribute to the pion yield.

This paper has focused on the published \sqsntwo\auau 0-20\% and 20-40\% direct photon \pt and $v_{2}$ distributions.  Future work to simulate the $q$-$\bar{q}$ photon contributions in more peripheral collisions is promising.  Additionally, the higher orders of the direct photon flow presents a new quantity to distinguish between the different photon processes.  Given the soft-gluon mediated $q$-$\bar{q}$ annihilation production mechanism ansatz, the $v_{n}$ for direct photons is expected to be similar to the pion $v_{n}$ at \pt less than 3~GeV/c for higher orders of $n$.  This model predicts that higher-order $v_{n}$ $n_{q}$-scaling laws seen with identified hadrons~\cite{ppg146} will remain valid for the direct photon $v_{n}$ where the $n_{q\gamma}=2$.

\begin{acknowledgments}

The author thanks John Lajoie and Paul Stankus for many valuable conversations.  Additional discussions with Bill Zajc, Peter Steinberg, Anne Sickles, Rich Petti, Volker Koch and Che-Ming Ko are also recognized as are the organizers and attendees at the 2013 ECT* workshop on "Electromagnetic Probes of Strongly Interacting Matter: Status and future of low-mass lepton-pair spectroscopy".  This research was supported by US Department of Energy grants DE-FG02-86ER40281 and DE-FG02-92ER40692.

\end{acknowledgments}

\bibliography{PRC_photonv2}

\begin{thebibliography}{29}
\expandafter\ifx\csname natexlab\endcsname\relax\def\natexlab#1{#1}\fi
\expandafter\ifx\csname bibnamefont\endcsname\relax
  \def\bibnamefont#1{#1}\fi
\expandafter\ifx\csname bibfnamefont\endcsname\relax
  \def\bibfnamefont#1{#1}\fi
\expandafter\ifx\csname citenamefont\endcsname\relax
  \def\citenamefont#1{#1}\fi
\expandafter\ifx\csname url\endcsname\relax
  \def\url#1{\texttt{#1}}\fi
\expandafter\ifx\csname urlprefix\endcsname\relax\def\urlprefix{URL }\fi
\providecommand{\bibinfo}[2]{#2}
\providecommand{\eprint}[2][]{\url{#2}}

\bibitem[{\citenamefont{Klein-Boesing}()}]{CKBThesis}
\bibinfo{author}{\bibfnamefont{C.}~\bibnamefont{Klein-Boesing}},
  \bibinfo{note}{ph.D. thesis, University of Muenster, 2005}.

\bibitem[{\citenamefont{Adare et~al.}(2010)}]{ppg86}
\bibinfo{author}{\bibfnamefont{A.}~\bibnamefont{Adare}} \bibnamefont{et~al.},
  \bibinfo{journal}{Physical Review Letters} \textbf{\bibinfo{volume}{104}},
  \bibinfo{pages}{132301} (\bibinfo{year}{2010}).

\bibitem[{\citenamefont{Adare et~al.}(2015)}]{ppg162}
\bibinfo{author}{\bibfnamefont{A.}~\bibnamefont{Adare}} \bibnamefont{et~al.},
  \bibinfo{journal}{Physical Review C} \textbf{\bibinfo{volume}{91}},
  \bibinfo{pages}{064904} (\bibinfo{year}{2015}).

\bibitem[{\citenamefont{Adare et~al.}(2012{\natexlab{a}})}]{ppg126}
\bibinfo{author}{\bibfnamefont{A.}~\bibnamefont{Adare}} \bibnamefont{et~al.},
  \bibinfo{journal}{Physical Review Letters} \textbf{\bibinfo{volume}{109}},
  \bibinfo{pages}{122302} (\bibinfo{year}{2012}{\natexlab{a}}).

\bibitem[{\citenamefont{for~the
  ALICE~Collaboration}(2013{\natexlab{a}})}]{Alicept}
\bibinfo{author}{\bibfnamefont{M.~W.} \bibnamefont{for~the
  ALICE~Collaboration}}, \bibinfo{journal}{Nuclear Physics A}
  \textbf{\bibinfo{volume}{904}}, \bibinfo{pages}{573c}
  (\bibinfo{year}{2013}{\natexlab{a}}).

\bibitem[{\citenamefont{for~the
  ALICE~Collaboration}(2013{\natexlab{b}})}]{Alicev2}
\bibinfo{author}{\bibfnamefont{D.~L.} \bibnamefont{for~the
  ALICE~Collaboration}}, \bibinfo{journal}{Journal of Physics:Conference
  Series} \textbf{\bibinfo{volume}{446}}, \bibinfo{pages}{012028}
  (\bibinfo{year}{2013}{\natexlab{b}}).

\bibitem[{\citenamefont{Chatterjee et~al.}(2006)\citenamefont{Chatterjee,
  Frodermann, Heinz, and Srivastava}}]{Heinz}
\bibinfo{author}{\bibfnamefont{R.}~\bibnamefont{Chatterjee}},
  \bibinfo{author}{\bibfnamefont{E.~S.} \bibnamefont{Frodermann}},
  \bibinfo{author}{\bibfnamefont{U.}~\bibnamefont{Heinz}}, \bibnamefont{and}
  \bibinfo{author}{\bibfnamefont{D.~K.} \bibnamefont{Srivastava}},
  \bibinfo{journal}{Physical Review Letters} \textbf{\bibinfo{volume}{96}},
  \bibinfo{pages}{202302} (\bibinfo{year}{2006}).

\bibitem[{\citenamefont{Chatterjee and Srivastava}(2009)}]{Chatterjee}
\bibinfo{author}{\bibfnamefont{R.}~\bibnamefont{Chatterjee}} \bibnamefont{and}
  \bibinfo{author}{\bibfnamefont{D.~K.} \bibnamefont{Srivastava}},
  \bibinfo{journal}{Physical Review C} \textbf{\bibinfo{volume}{79}},
  \bibinfo{pages}{021901} (\bibinfo{year}{2009}).

\bibitem[{\citenamefont{Liu et~al.}(2009)\citenamefont{Liu, Hirano, Werner, and
  Zhu}}]{Liu}
\bibinfo{author}{\bibfnamefont{F.-M.} \bibnamefont{Liu}},
  \bibinfo{author}{\bibfnamefont{T.}~\bibnamefont{Hirano}},
  \bibinfo{author}{\bibfnamefont{K.}~\bibnamefont{Werner}}, \bibnamefont{and}
  \bibinfo{author}{\bibfnamefont{Y.}~\bibnamefont{Zhu}},
  \bibinfo{journal}{Physical Review C} \textbf{\bibinfo{volume}{80}},
  \bibinfo{pages}{034905} (\bibinfo{year}{2009}).

\bibitem[{\citenamefont{Liu and Liu}(2014)}]{Liu2}
\bibinfo{author}{\bibfnamefont{F.-M.} \bibnamefont{Liu}} \bibnamefont{and}
  \bibinfo{author}{\bibfnamefont{S.-X.} \bibnamefont{Liu}},
  \bibinfo{journal}{Physical Review C} \textbf{\bibinfo{volume}{89}},
  \bibinfo{pages}{034906} (\bibinfo{year}{2014}).

\bibitem[{\citenamefont{Muller et~al.}()\citenamefont{Muller, Wu, and
  Yang}}]{Muller}
\bibinfo{author}{\bibfnamefont{B.}~\bibnamefont{Muller}},
  \bibinfo{author}{\bibfnamefont{S.-Y.} \bibnamefont{Wu}}, \bibnamefont{and}
  \bibinfo{author}{\bibfnamefont{D.-L.} \bibnamefont{Yang}},
  \bibinfo{note}{arXiv:1308.6568}.

\bibitem[{\citenamefont{Tuchin}(2013)}]{Tuchin}
\bibinfo{author}{\bibfnamefont{K.}~\bibnamefont{Tuchin}},
  \bibinfo{journal}{Physical Review C} \textbf{\bibinfo{volume}{87}},
  \bibinfo{pages}{024912} (\bibinfo{year}{2013}).

\bibitem[{\citenamefont{Chiu et~al.}(2013)\citenamefont{Chiu, Hemmick,
  Khachatryan, Leonidov, Liao, and McLerran}}]{Glasma}
\bibinfo{author}{\bibfnamefont{M.}~\bibnamefont{Chiu}},
  \bibinfo{author}{\bibfnamefont{T.~K.} \bibnamefont{Hemmick}},
  \bibinfo{author}{\bibfnamefont{V.}~\bibnamefont{Khachatryan}},
  \bibinfo{author}{\bibfnamefont{A.}~\bibnamefont{Leonidov}},
  \bibinfo{author}{\bibfnamefont{J.}~\bibnamefont{Liao}}, \bibnamefont{and}
  \bibinfo{author}{\bibfnamefont{L.}~\bibnamefont{McLerran}},
  \bibinfo{journal}{Nuclear Physics A} \textbf{\bibinfo{volume}{900}},
  \bibinfo{pages}{16} (\bibinfo{year}{2013}).

\bibitem[{\citenamefont{van Hees et~al.}(2011)\citenamefont{van Hees, Gale, and
  Rapp}}]{Rapp}
\bibinfo{author}{\bibfnamefont{H.}~\bibnamefont{van Hees}},
  \bibinfo{author}{\bibfnamefont{C.}~\bibnamefont{Gale}}, \bibnamefont{and}
  \bibinfo{author}{\bibfnamefont{R.}~\bibnamefont{Rapp}},
  \bibinfo{journal}{Physical Review C} \textbf{\bibinfo{volume}{84}},
  \bibinfo{pages}{054906} (\bibinfo{year}{2011}).

\bibitem[{\citenamefont{Linnyk et~al.}(2014)\citenamefont{Linnyk, Cassing, and
  Bratkovskaya}}]{Linnyk}
\bibinfo{author}{\bibfnamefont{O.}~\bibnamefont{Linnyk}},
  \bibinfo{author}{\bibfnamefont{W.}~\bibnamefont{Cassing}}, \bibnamefont{and}
  \bibinfo{author}{\bibfnamefont{E.~L.} \bibnamefont{Bratkovskaya}},
  \bibinfo{journal}{Physical Review C} \textbf{\bibinfo{volume}{89}},
  \bibinfo{pages}{034908} (\bibinfo{year}{2014}).

\bibitem[{\citenamefont{Adler et~al.}(2003)}]{hadronflow}
\bibinfo{author}{\bibfnamefont{S.~S.} \bibnamefont{Adler}}
  \bibnamefont{et~al.}, \bibinfo{journal}{Physical Review Letters}
  \textbf{\bibinfo{volume}{91}}, \bibinfo{pages}{182301}
  (\bibinfo{year}{2003}).

\bibitem[{\citenamefont{Molnar and Voloshin}(2003)}]{nqscale}
\bibinfo{author}{\bibfnamefont{D.}~\bibnamefont{Molnar}} \bibnamefont{and}
  \bibinfo{author}{\bibfnamefont{S.~A.} \bibnamefont{Voloshin}},
  \bibinfo{journal}{Physical Review Letters} \textbf{\bibinfo{volume}{91}},
  \bibinfo{pages}{092301} (\bibinfo{year}{2003}).

\bibitem[{\citenamefont{Greco et~al.}(2003)\citenamefont{Greco, Ko, and
  Levai}}]{coalescence}
\bibinfo{author}{\bibfnamefont{V.}~\bibnamefont{Greco}},
  \bibinfo{author}{\bibfnamefont{C.~M.} \bibnamefont{Ko}}, \bibnamefont{and}
  \bibinfo{author}{\bibfnamefont{P.}~\bibnamefont{Levai}},
  \bibinfo{journal}{Physical Review C} \textbf{\bibinfo{volume}{68}},
  \bibinfo{pages}{034904} (\bibinfo{year}{2003}).

\bibitem[{\citenamefont{Adare et~al.}({\natexlab{a}})}]{ppg146}
\bibinfo{author}{\bibfnamefont{A.}~\bibnamefont{Adare}} \bibnamefont{et~al.},
  \bibinfo{note}{arXiv:1412.1038}.

\bibitem[{\citenamefont{Adare et~al.}(2012{\natexlab{b}})}]{ppg123}
\bibinfo{author}{\bibfnamefont{A.}~\bibnamefont{Adare}} \bibnamefont{et~al.},
  \bibinfo{journal}{Physical Review C} \textbf{\bibinfo{volume}{85}},
  \bibinfo{pages}{064914} (\bibinfo{year}{2012}{\natexlab{b}}).

\bibitem[{\citenamefont{Adare et~al.}(2007)}]{ppg61}
\bibinfo{author}{\bibfnamefont{A.}~\bibnamefont{Adare}} \bibnamefont{et~al.},
  \bibinfo{journal}{Physical Review Letters} \textbf{\bibinfo{volume}{98}},
  \bibinfo{pages}{162301} (\bibinfo{year}{2007}).

\bibitem[{\citenamefont{Adare et~al.}({\natexlab{b}})}]{ppg124}
\bibinfo{author}{\bibfnamefont{A.}~\bibnamefont{Adare}} \bibnamefont{et~al.},
  \bibinfo{note}{arXiv:1412:1043}.

\bibitem[{\citenamefont{Schnedermann et~al.}(1993)\citenamefont{Schnedermann,
  Sollfrank, and Heinz}}]{BlastWave}
\bibinfo{author}{\bibfnamefont{E.}~\bibnamefont{Schnedermann}},
  \bibinfo{author}{\bibfnamefont{J.}~\bibnamefont{Sollfrank}},
  \bibnamefont{and} \bibinfo{author}{\bibfnamefont{U.}~\bibnamefont{Heinz}},
  \bibinfo{journal}{Physical Review C} \textbf{\bibinfo{volume}{48}},
  \bibinfo{pages}{2462} (\bibinfo{year}{1993}).

\bibitem[{\citenamefont{Tang et~al.}(2009)\citenamefont{Tang, Xu, Ruan, van
  Buren, Wang, and Xu}}]{Lijuan}
\bibinfo{author}{\bibfnamefont{Z.}~\bibnamefont{Tang}},
  \bibinfo{author}{\bibfnamefont{Y.}~\bibnamefont{Xu}},
  \bibinfo{author}{\bibfnamefont{L.}~\bibnamefont{Ruan}},
  \bibinfo{author}{\bibfnamefont{G.}~\bibnamefont{van Buren}},
  \bibinfo{author}{\bibfnamefont{F.}~\bibnamefont{Wang}}, \bibnamefont{and}
  \bibinfo{author}{\bibfnamefont{Z.}~\bibnamefont{Xu}},
  \bibinfo{journal}{Physical Review C} \textbf{\bibinfo{volume}{79}},
  \bibinfo{pages}{051901(R)} (\bibinfo{year}{2009}).

\bibitem[{\citenamefont{Retiere and Lisa}(2004)}]{Lisa}
\bibinfo{author}{\bibfnamefont{F.}~\bibnamefont{Retiere}} \bibnamefont{and}
  \bibinfo{author}{\bibfnamefont{M.~A.} \bibnamefont{Lisa}},
  \bibinfo{journal}{Physical Review C} \textbf{\bibinfo{volume}{70}},
  \bibinfo{pages}{044907} (\bibinfo{year}{2004}).

\bibitem[{\citenamefont{Hwa and Yang}(2004)}]{Hwa}
\bibinfo{author}{\bibfnamefont{R.~C.} \bibnamefont{Hwa}} \bibnamefont{and}
  \bibinfo{author}{\bibfnamefont{C.~B.} \bibnamefont{Yang}},
  \bibinfo{journal}{Physical Review C} \textbf{\bibinfo{volume}{70}},
  \bibinfo{pages}{024905} (\bibinfo{year}{2004}).

\bibitem[{\citenamefont{Afanasiev et~al.}(2012)}]{ppg139}
\bibinfo{author}{\bibfnamefont{S.}~\bibnamefont{Afanasiev}}
  \bibnamefont{et~al.}, \bibinfo{journal}{Physical Review Letters}
  \textbf{\bibinfo{volume}{109}}, \bibinfo{pages}{152302}
  (\bibinfo{year}{2012}).

\bibitem[{\citenamefont{Adare et~al.}(2012{\natexlab{c}})}]{ppg136}
\bibinfo{author}{\bibfnamefont{A.}~\bibnamefont{Adare}} \bibnamefont{et~al.},
  \bibinfo{journal}{Physical Review D} \textbf{\bibinfo{volume}{86}},
  \bibinfo{pages}{072008} (\bibinfo{year}{2012}{\natexlab{c}}).

\bibitem[{\citenamefont{Adler et~al.}(2007)}]{ppg60}
\bibinfo{author}{\bibfnamefont{S.~S.} \bibnamefont{Adler}}
  \bibnamefont{et~al.}, \bibinfo{journal}{Physical Review Letters}
  \textbf{\bibinfo{volume}{98}}, \bibinfo{pages}{012002}
  (\bibinfo{year}{2007}).

\end{thebibliography}

\end{document}